\documentclass[prl,twocolumn,showpacs,preprintnumbers,amsmath,amssymb,superscriptaddress]{revtex4}
\usepackage{graphicx}
\begin{document}

\title{Peculiar behavior of the electrical resistivity of MnSi at the ferromagnetic phase transition}

\author{Alla E. Petrova}
\affiliation{Institute for High Pressure Physics of Russian
Academy of Sciences, Troitsk, Moscow Region, Russia}
\author{Eric Bauer}
\affiliation{Los Alamos National Laboratory, Los Alamos, 87545 NM,
USA}
\author{Vladimir Krasnorussky}
\affiliation{Institute for High Pressure Physics of Russian
Academy of Sciences, Troitsk, Moscow Region, Russia}
\author{Sergei M. Stishov}
\email{sergei@hppi.troitsk.ru} \affiliation{Institute for High
Pressure Physics of Russian Academy of Sciences, Troitsk, Moscow
Region, Russia} \affiliation{Los Alamos National Laboratory, Los
Alamos, 87545 NM, USA}

\date{\today}

\begin{abstract}
The electrical resistivity of a single crystal of MnSi was
measured across its ferromagnetic phase transition line at ambient
and high pressures. Sharp peaks of the temperature coefficient of
resistivity characterize the transition line. Analysis of these
data shows that at pressures to $\sim0.35$ GPa these peaks have
fine structure, revealing a shoulder at $\sim0.5$ K above the
peak. It is symptomatic that this structure disappears at
pressures higher than $\sim0.35$ GPa, which was identified earlier
as a tricritical point.
\end{abstract}


\maketitle

The intermetallic compound MnSi experiences a second order phase
transition at temperature $T_{c}$ slightly below 30 K, acquiring
helical magnetic structure and becoming a weak itinerant
ferromagnet. On application of pressure the transition temperature
$T_{c}$ decreases and tends to zero at a pressure of about 1.4 GPa
\cite{1}. As was noticed for the first time in ref. \cite{2} (see
also \cite{3}), a $\lambda$-type singularity of the AC magnetic
susceptibility $\chi_{AC}$ at the phase transition in MnSi deforms
gradually with pressure and becomes a simple step at pressures
more than 1GPa. That was ground to claim existence of a
tricritical point with the coordinates: $\sim1.2$ GPa, $\sim12$ K
\cite{2,3}. This conclusion was partly disputed in ref. \cite{4},
where new measurements of $\chi_{AC}$ of MnSi at high pressures,
created by compressed helium, were reported. These authors
\cite{4} confirmed the existence of a tricritical point on the
phase transition line in MnSi but placed it at much lower pressure
and at significantly higher temperature ($P_{tr}\cong0.355$ GPa,
$T_{tr}\cong25.2$ K).

To resolve that somewhat controversial issue we have carried out
precise resistivity measurements of a MnSi single crystal across
the phase transition line at ambient and at high pressures, using
a compressed helium technique. The primary purpose was to study
behavior of the temperature coefficient of resistivity $d\rho/dT$
at the transition line. According to the theoretical conclusions
\cite{5,6,7}, a temperature coefficient of resistivity diverge at
a second order magnetic phase transition and can be characterized
by a static critical exponent . Contrary to our expectations, we
found that peaks in $d\rho/dT$ at $T_{c}$ are accompanied by a
well-defined shoulder on their high temperature side, which vanish
when approaching a pressure $\sim0.35$ GPa. This finding nicely
correlates with corresponding features in ultrasound attenuation
\cite{8}, thermal expansion \cite{9}, and heat capacity \cite{10},
discovered in the critical region of MnSi at ambient pressure.

The single crystal of MnSi was grown from a tin flux by dissolving
pre-alloyed Mn and Si in excess Sn. For resistivity measurements,
four Pt-wires of 25-$\mu$m in diameter were welded to the crystal
with dimensions of about $0.5{\times}0.3{\times}0.3 mm^{3}$. The
temperature of the magnetic phase transition $T_{c}$ and
resistivity ratio $R_{300}/R_{(T=2.1)}$, taken at ambient
pressure, are equal to $29.25{\pm}0.02$ K and $\approx100$
correspondingly. The crystal was placed into a high pressure cell
made of beryllium copper. Fluid and solid helium were used as a
pressure medium. Temperature was measured by a calibrated Cernox
sensor, imbedded in the cell body, with an accuracy of about 0.05
K. A calibrated manganin gauge was used to measure pressure with
accuracy about $10^{-3}$ GPa in the fluid helium domain. In the
domain of solid helium, pressure was calculated on the basis of
the measured helium-crystallization temperatures and data for the
equation of state of helium. Accuracy of pressure measurements in
solid helium is estimated as $5{\times}10^{-3}$ GPa. The
resistivity was measured by a four-terminal DC method. The
experimental setup, including the high pressure gas installation
and the cryostat, is described in \cite{4}, \cite{11}.

\begin{figure}[htb]
\includegraphics[width=80mm]{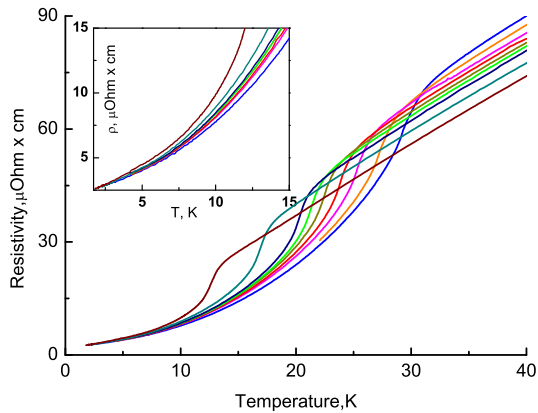}
\caption{\label{fig1} (Color online) Temperature dependence of the
electrical resistivity ${\rho(T)}$ of MnSi at different pressures.
The isobars correspond to pressures in GPa: 0, 0.2, 0.32, 0.43,
0.54, 0.63, 0.7, 0.885, 1.13, counting from the right to the left
at the bottom of the figure. }
\end{figure}

The resistivity measurements of MnSi were carried out along 24
quasi isobars \cite{Iso} in the pressure range from zero to 1.5
GPa. Selected experimental data are displayed in Fig.\ref{fig1}.
We have tried to describe the resistivity curves in the
temperature range from zero to the phase transition region by
various polynomials that contained potentially important $T^{2}$
and/or $T^{5}$ terms accounting for scattering by spin and density
fluctuations (phonons) \cite{12, 13}. The overall results appeared
to be quite satisfactory though we observed small but systematic
deviations of the experimental data points from the corresponding
approximations at low temperatures. Replacing the $T^{2}$ term
with $T^{n}$ improves the situation but does not correct it
entirely, though always leads to a value of  $n<2$. On the other
hand, as is seen in Fig.\ref{fig1} the pressure derivatives of
resistivity are positive below the Curie point and negative above
(see also \cite{3}). This implies a dominant role of order
parameter fluctuations in the electron scattering in MnSi. Hence,
any analysis of the resistivity behavior in MnSi should take into
account this significant factor. We will discuss this issue
elsewhere. However, it is important to emphasize here that the
residual resistivity of MnSi, derived from reasonable
extrapolations, decreases monotonically from 2.25 to 2.11
$\mu\Omega$cm over all the pressure range studied on compression.
This indicates that many cycles of pressure loading and unloading,
cooling and warming do not introduce additional defects into the
sample. The temperature-dependent resistivity of MnSi above the
phase transition line shows clear signs of resistivity saturation
at $T\rightarrow\propto$ \cite{14}.

\begin{figure}[htb]
\includegraphics[width=80mm]{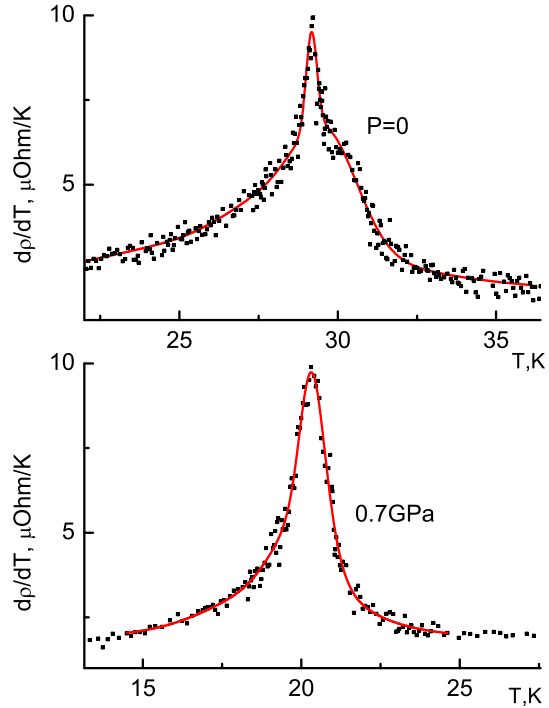}
\caption{\label{fig2} (Color online) Examples of temperature
derivatives of resistivity $d\rho/dT$ at ambient and elevated
pressures. The square dots are the temperature derivatives of
resistivity, taken by averaging the slopes of two adjacent points
of the raw experimental data. The solid lines are results of
smoothing procedures. }
\end{figure}

\begin{figure}[htb]
\includegraphics[width=80mm]{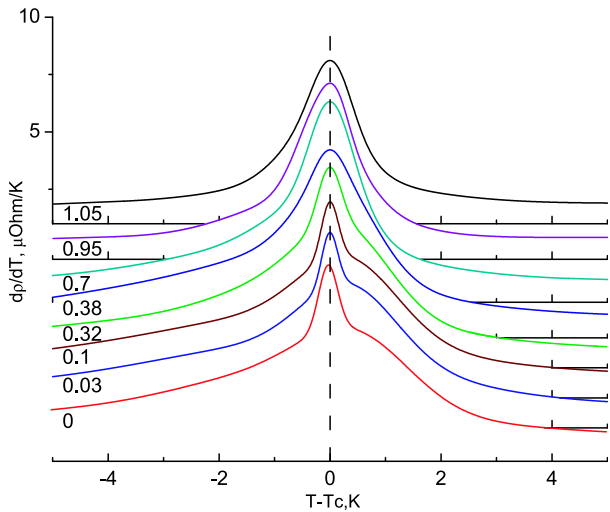}
\caption{\label{fig3} (Color online) Evolution of temperature
derivatives of resistivity $d\rho/dT$ with pressure. The pressures
in GPa are shown at the left side of the figure. }
\end{figure}

Now we turn to an analysis of the temperature coefficient of
resistivity $d\rho/dT$ in the vicinity of the phase transition
boundary. Temperature derivatives of resistivity $\rho$ were taken
by averaging the slopes of two adjacent points of the raw
experimental data. The outcome of this procedure is illustrated in
Fig.\ref{fig2}, where also the smoothing lines are shown. As is
seen from the figure at ambient pressure the curve ${d\rho\over
dT}(T)$ has a distinct shoulder on the high temperature side of
$T_{c}$ which disappears at high pressure. The evolution of the
shape of the peaks of $d\rho/dT$ with applied pressure is shown in
Fig.\ref{fig3}. The shoulder in ${d\rho\over dT}(T)$ vanishes at a
pressure of around 0.35 GPa that was recognized early as a
coordinate of the tricritical point \cite{4}. The overall trend is
that at low pressure structure in $d\rho/dT$ consists of two
components: one sharp and another broad, separated only by half a
degree or so. Because of lack of \textit{a priori} knowledge of
the peak forms and uncertainty with background subtraction, we
could not separate these peaks in a reliable way. The obvious
overlapping of the peaks makes also unreliable attempts to obtain
a critical exponent, based on behavior $d\rho/dT$ \cite{5,6,7}.
Nevertheless, we have found that an approximation of $d\rho/dT$ at
$T<T_{c}$ with the expression
\begin{equation}
 {d\rho \over dT}=a+bT+c(T_{c}-T)^{-m}
\end{equation}
gives $m\approx0.25$ in case of the low pressure isobars, which is
a reasonable value for an exponent characterizing critical
behavior of heat capacity near helical spin ordering \cite{15,
16}. At pressures more than 0.3-0.4 GPa, the fitting became
unstable and did not lead to realistic values of the exponents.
\begin{figure}[htb]
\includegraphics[width=80mm]{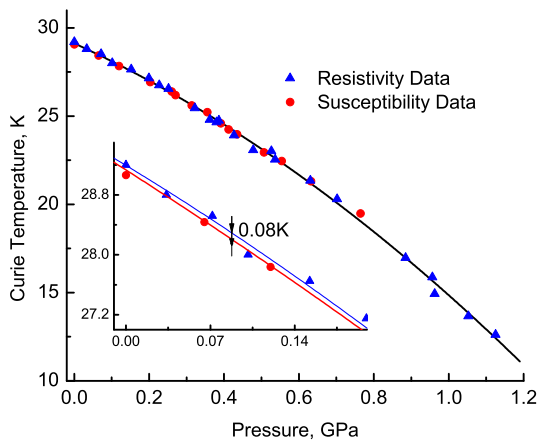}
\caption{\label{fig4} (Color online) Pressure dependence of the
Curie temperature of MnSi according to the current resistivity
measurements and the AC susceptibility data \cite{4}. The inset
shows that the average mismatch of the two sets of the data is
less than 0.1K. }
\end{figure}

Summarizing, we point out that the reported experimental data
demonstrate complicated behavior of the temperature coefficient of
resistivity of MnSi in the vicinity of its phase transition. As is
seen from Fig.\ref{fig3}, $d\rho/dT$ evolves from a highly
asymmetric, not quite resolved doublet with one rather sharp
component at ambient pressure to the single, fairly symmetric peak
at pressure, corresponding to the tricritical point. It was
mentioned earlier that a doublet structure of related peaks was
discovered in sound absorption \cite{8}, thermal expansion
\cite{9}, and heat capacity \cite{10} at the phase transition in
MnSi that correlate with the current observations. Unfortunately,
little is known about the origin of this structure, but what we
know is that the high temperature satellite does not reveal itself
in magnetic susceptibility measurements \cite{3,4}. The data
comparison shows that the magnetic transition is associated with a
sharp peak on the low temperature side of $d\rho/dT$
(Fig.\ref{fig4}). Thus, the observed shoulder in $d\rho/dT$ could
be connected with short range spin order or with the spin texture
\cite{18, 19}. However, it does not appear that the shoulder in
$d\rho/dT$ marks any kind of a conventional phase transition.
Nevertheless, one cannot exclude that a topological phase
transition takes place at a temperature above the magnetic
transformation. In the latter case, instead of a tricritical point
there would be a special kind of a multicritical point in the
phase diagram of MnSi. But, if the scenario with a topological
phase transition is not appropriate, then the shoulder in
$d\rho/dT$ disappears, being adsorbed by the volume instability
gap, which is opened at a tricritical point \cite{20}.
\begin{acknowledgments}
Authors express their gratitude to Vladimir Sidorov for technical
assistance and to J. D. Thompson for reading the manuscript and
valuable remarks. A.E. Petrova, V. Krasnorussky and S.M. Stishov
appreciate support of the Russian foundation for Basic Research
(grant 06-02-16590), Program of the Physics Department of Russian
Academy of Science on Strongly Correlated Systems and Program of
the Presidium of Russian Academy of Science on Physics of Strongly
Compressed Matter. Work at Los Alamos was performed under the
auspices of the US Department of Energy, Office of Science.
\end{acknowledgments}


\begin{thebibliography}{99}
\bibitem{Iso} The normal experimental procedure starts with cooling the high pressure cell, containing certain amount of compressed helium and the sample. After a while helium crystallizes, blocking the high pressure tubing [4,11] and making further cooling isochoric in respect to solid helium. So cooling and subsequently warming the sample of MnSi, which has compressibility quite different from that of helium, are neither isobaric nor isochoric.
\bibitem{20} At a first order phase transition, a volume change occurs since a homogenous state of matter becomes instable in a certain range of volumes, which may be called a volume gap. Specific features of behavior around a second order phase transition could fall in the volume gap when the transition turns to be a first order and therefore may not be observable.
\end{thebibliography}

\end{document}